\documentstyle[procl, twoside]{article}
\input{psfig.sty}
\def\Journal#1#2#3#4{(#1) {#2} {\bf #3}, #4}

\def\ApJ{\em Astrophys.~J.}
\def\ApJL{\em Astrophys.~J., Lett.}

\def\AaAp{\em Astron. Astrophys.}

\def\MNRAS{\em Mon. Not. R.~Astron. Soc.}
\def\Nat{\em Nature\/}

\newcommand{\HI}{{\rm H\,\scriptstyle I}}
\newcommand{\HII}{{\rm H\,\scriptstyle II}}
\begin{document}
\markboth{Christine D. Wilson}{Molecular Gas in M33}
\thispagestyle{plain}

\title{Physical Conditions and Spatial Structure in the Molecular Interstellar Medium of M33}
\author{Christine D. Wilson}
\address{Dept. of Physics and Astronomy, McMaster University\\
Hamilton, Ontario L8S 4M1 Canada}

\maketitle

\abstract{Early studies of molecular gas in M33 used the CO J=1-0
transition to map out the total gas content in the inner kiloparsec of
the galaxy and to study the properties of the giant molecular cloud
population. In this review, 
I discuss recent detailed studies of individual molecular clouds
in M33 in higher rotational transitions of CO, $^{13}$CO, and the
fine structure line of atomic carbon. These data reveal that the
temperature and density of the molecular gas is correlated with the
presence of nearby massive star formation, with clouds with more 
intense star formation having higher temperatures and lower densities.
This effect is most likely due to post-star formation processing of
the molecular gas. A comparison of $^{13}$CO observations for M33
and individual molecular clouds in the Milky Way suggests that a significant
fraction of the molecular interstellar medium resides in low column
density regions and accounts for 30-60\% of the CO luminosity. Finally,
a detailed map of a single molecular cloud in the giant $\HII$ region
NGC 604 reveals an offset between the atomic carbon and CO emission,
which indicates that the dominant source of atomic carbon in this cloud
is likely photodissociation of CO by ultraviolet 
photons from the massive stars.
A complete map of the CO emission in M33 would be extremely helpful to
guide further detailed studies.
}

\section{Introduction}

Although M33 is substantially smaller than M31, the total H$\alpha$ 
luminosities
and hence global star formation rates of the two galaxies are quite similar.
This higher star formation rate per unit area, along with the presence
of very luminous $\HII$ regions such as NGC 604 and NGC 595, makes M33 an 
interesting target for studying its molecular interstellar medium.
The first observations of CO emission in M33 (Young and Scoville 1982;
Blitz 1985) suggested that the molecular gas content of this galaxy was
relatively low. However, a complete map of the inner kiloparsec of M33
revealed significant CO emission (Wilson and Scoville 1989) 
corresponding to $3\times 10^7$ M$_\odot$ of molecular gas. Interferometric
observations (Wilson et al. 1988; Wilson and Scoville 1990, 1992; Viallefond
et al. 1992) have revealed a population of giant molecular clouds with 
sizes, line widths, and masses that are very similar to molecular clouds
in the Milky Way. A recent reanalysis of the M33 data along with similar
data sets for M31 and the Milky Way by Sheth et al. (these proceedings) 
uses several different techniques for
identifying and analyzing the molecular clouds. The results of that 
study show that the properties of molecular
clouds in these three galaxies are very similar when the clouds are all
analyzed in a self-consistent way.

Although there have been major campaigns over
the last 10 years to map the total molecular disk of M31 (Dame et al. 1993;
Neininger et al. 1998; Loinard et al. 1999; 
Gu\'elin, these proceedings), we still do
not have a complete map of the CO emission in M33. Therefore,
instead of discussing the global properties of the molecular
interstellar medium in M33, I will 
focus on recent detailed analyses of specific regions in M33. These
studies use
observations of the higher rotational transitions of $^{12}$CO, the
rarer isotopomer $^{13}$CO, and the fine structure line of atomic carbon to
study the physical conditions and structure of the molecular 
interstellar medium in M33. In Section 2, I discuss how observations
of multiple rotational transitions of $^{12}$CO and
$^{13}$CO can be used to constrain the
density and temperature of the molecular gas, and how these properties
relate to the local star formation environment. In Section 3, I review the
evidence for a component of molecular gas at low column densities from
observations of $^{13}$CO in M33 and in molecular clouds in the Milky Way.
In Section 4, I summarize recent observations of atomic carbon in
several molecular clouds in M33 and, in particular, a map of
the [CI] emission from a molecular cloud in the giant $\HII$ region NGC 604. 
I conclude with some suggestions for future work on the molecular
interstellar medium in M33.

\section{Physical Properties of Molecular Clouds and the Link to 
Star Formation}

It seems reasonable to expect that the star formation properties of a molecular
cloud may be affected by the physical conditions in the cloud. For
example, it is sometimes suggested that the types of stars that form 
may be affected by the temperature of the cloud (Turner 1984), 
while the total
star formation efficiency of a cloud may be tied to the mass fraction
of the cloud in dense cores (i.e. Lada et al. 1991). 
However, studying the link between
star formation and molecular cloud properties is complicated by the fact
that star formation can in turn alter the physical conditions in the
molecular cloud by heating, compressing, or photodissociating the molecular
gas. The key question here is which process is dominant: are differences in
the star formation properties among clouds due to the current
physical conditions of the clouds, or do those physical conditions instead
reflect primarily post-star formation processing of the molecular gas?
M33 offers an excellent opportunity to try to disentangle these effects,
since we can study individual molecular clouds in a variety of star formation
environments all at a common distance.

\vskip15pt

\begin{figure}[htb]
\centerline{\psfig{figure=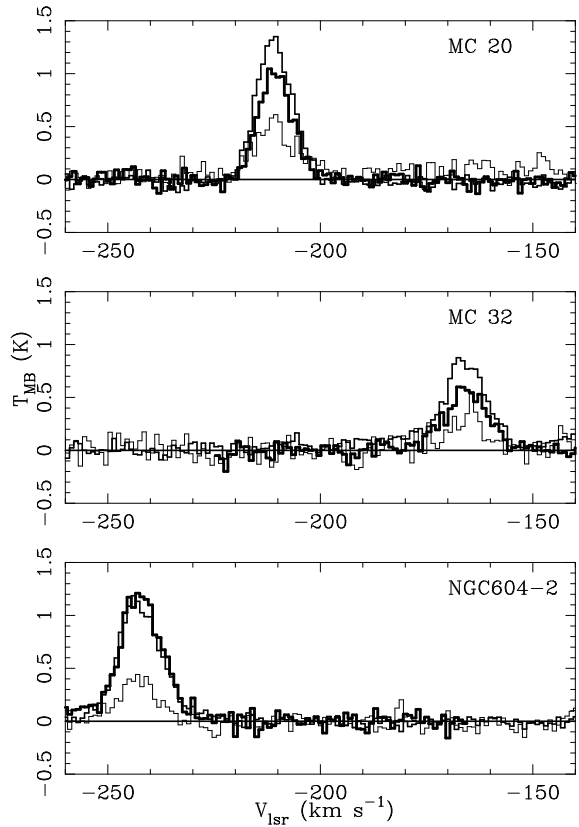,width=8truecm,%
       bbllx=32pt,bblly=84pt,bburx=564pt,bbury=709pt}}
\vskip15pt
\vskip15pt
\caption{$^{12}$CO J=3-2 (thick line), $^{12}$CO J=2-1 (medium line),
and $^{13}$CO J=2-1 (thin line, scaled up by a factor of 3) for three
giant molecular clouds in M33 (Wilson et al. 1997). }
\end{figure}

Measuring the density and temperature of extragalactic molecular clouds
is made more difficult by the sizes and internal structure of
the clouds themselves. Molecular clouds are clumpy, which means that
the typical density of the material producing the CO emission 
does not equal the volume-averaged
density ($\overline\rho = M/V$). 
In addition, single molecular clouds are not large enough to
fill the beam of even large single radio
telescopes, which means that the kinetic temperature of the cloud is not
equal to  the observed antenna temperature. The solution to
these problems is to combine observations of several rotational transitions
of $^{12}$CO and $^{13}$CO with radiative transfer models to constrain
the density and temperature of the molecular gas. 
Wilson, Walker, and Thornley (1997) combined observations of three 
lines of $^{12}$CO and two lines of $^{13}$CO
with a large velocity gradient analysis to obtain individual estimates
of the density and temperature for seven giant molecular clouds in M33.
Figure~1 shows some of the lines observed for three of the clouds;
note in particular how the strength of the $^{12}$CO J=3-2 line increases
relative to the $^{12}$CO J=2-1 line as we move from a cloud without
an optical $\HII$ region (MC32), to a cloud with a normal $\HII$ region
(MC20), to a cloud located very near a giant $\HII$ region (NGC 604-2).
Although the $^{12}$CO/$^{13}$CO J=1-0 and J=2-1 line ratios are observed
to be very uniform throughout the cloud sample, the $^{12}$CO J=3-2/2-1
line ratios increase by a factor of almost two in the cloud near the
giant $\HII$ region compared to clouds without optically visible
star formation.

The density and temperature of the molecular gas derived from the 
large velocity gradient models also show significant changes 
that are correlated with the star formation intensity (Table~1).
In particular, the molecular gas located near the giant $\HII$ region
NGC 604 is both hotter and less dense than the molecular gas
in more quiescent regions. In addition, the volume filling factor of
the dense gas (which is equal to the ratio of the volume-averaged
density to the density derived from the radiative transfer model) is
highest near the giant HII region and decreases towards more quiescent regions.
Are the unusual physical conditions seen in this cloud likely to be
the reason why this region has formed a giant $\HII$ region, or do
the physical conditions instead reflect the local star formation 
environment? The intense ultraviolet radiation field produced by the
massive stars is likely to increase the kinetic temperature of the
molecular gas. The expected effect of star formation on density is
a little less clear; density could be increased through shocks and
increases in the ambient pressure, but might be decreased through
photo-dissociation of the molecules and through conversion of the densest
regions into young stars. However, given that a 
continuous range of all physical conditions (temperature, density,
and filling factor) is observed as we move from clouds without optical
$\HII$ regions to clouds with normal $\HII$ regions to a cloud in a giant
$\HII$ region, it seems more likely that the different physical conditions
in these clouds are due to post-star formation changes in the molecular gas,
rather than the intrinsic conditions of the pre-star formation molecular cloud.

\begin{table}[htb]
\caption{Physical Conditions in M33 Clouds (from Wilson et al. 1997)}
\begin{center}
\begin{tabular}{llll} \noalign{\medskip}
\hline\noalign{\smallskip}
Clouds & $T_k$  & $n(H_2)$  & filling factor  \\
\noalign{\smallskip}
\hline\noalign{\smallskip}
NGC604-2 & $\ge 100$ K & $(1-3)\times 10^3$ cm$^{-3}$   & 17-50\% \\
\noalign{\smallskip}
with $\HII$ regions & $15-100$ K & $(2-10)\times 10^3$ cm$^{-3}$   & 2-20\% \\
\noalign{\smallskip}
without $\HII$ regions & $10-20$ K & $(5-30)\times 10^3$ cm$^{-3}$   & 0.1-4\% \\
\noalign{\smallskip}
\hline
\noalign{\smallskip}
\end{tabular}
\end{center}
\end{table}

\section{The Structure of the Molecular Interstellar Medium}

Comparing observations of molecular emission lines in galaxies like
M33 with similar data for molecular clouds in the Milky Way can provide
information on the large-scale structure of the molecular interstellar
medium. For example, 
early observations of the $^{13}$CO emission from external galaxies
revealed that the
$^{12}$CO/$^{13}$CO line ratio is significantly larger in the disks
of spiral galaxies than it is in giant molecular clouds in our own Galaxy
(Encrenaz et al. 1979), a difference which has persisted as the galaxy
sample has grown. Possible explanations for this difference include
a different filling factor in the two lines in the large extragalactic beams
(Encrenaz et al. 1979), different [$^{12}$CO]/[$^{13}$CO] abundance
ratios from one galaxy to another (Casoli, Dupraz, and Combes 1992),
and a mixture of giant molecular clouds and low column density (or ``diffuse'')
molecular material in galactic disks (Polk et al. 1988). 
M33 presents us with a unique
opportunity to test these different scenarios, since we can observe both
individual molecular clouds using millimeter interferometers, and larger
areas of the disk using single dish telescopes.

Wilson and Walker (1994) detected $^{13}$CO J=1-0 emission towards
eight giant molecular clouds in M33 using the NRAO 12~m telescope.
The average $^{12}$CO/$^{13}$CO intensity ratio measured 
in the 55$^{\prime\prime}$ beam is $10.0\pm0.09$, which is roughly a factor of two larger than the value observed in the Milky Way. This sample shows
no significant variation in the line ratio over regions that vary in
metallicity by a factor of three, which rules out metallicity as 
a factor in the different line ratios, at least for M33.
They also mapped a single molecular
cloud interferometrically in the $^{13}$CO line using the Caltech
Millimeter Array. The molecular cloud has the same filling factor in
both $^{12}$CO and $^{13}$CO (Figure 2), and thus different filling factors
in the different molecules do not seem to be an important factor in this 
region.
The intensity ratio obtained from the interferometer map
is $7.5\pm2.1$, which is consistent (within the rather large error bars)
with the values of 3-6 obtained in the Milky Way. 
Thus, the high line ratio observed over 200 pc regions
of the disk of M33 seems most likely to be produced by a mixture of
low and high column density gas within the radio beam. Wilson and Walker
(1994) estimate that between 30\% and 60\% of the $^{12}$CO intensity
in these regions originates in gas with a lower average column density.
A significant contribution from low column density gas would affect
estimates of the total mass of molecular gas in these regions, since
using the standard CO-to-H$_2$ conversion factor for this material would
produce an overestimate of the total gas mass.

\vskip15pt
\begin{figure}[htb]
\centerline{\psfig{figure=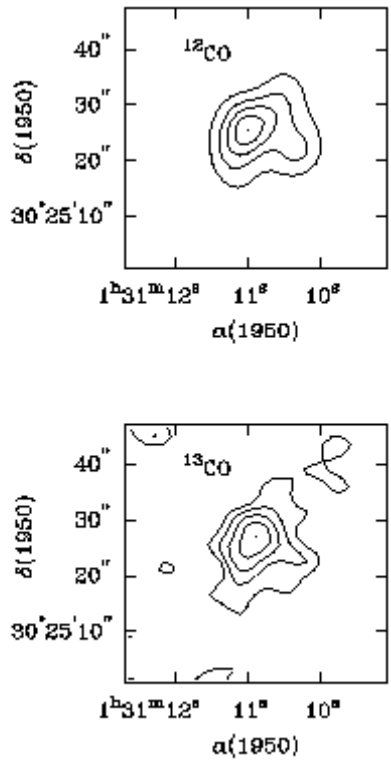,width=8truecm,%
       bbllx=32pt,bblly=84pt,bburx=564pt,bbury=709pt}}
\caption{$^{12}$CO and $^{13}$CO images of the molecular cloud MC20 in M33
(Wilson and Walker 1994). }
\end{figure}

\begin{figure}[htb]
\centerline{\psfig{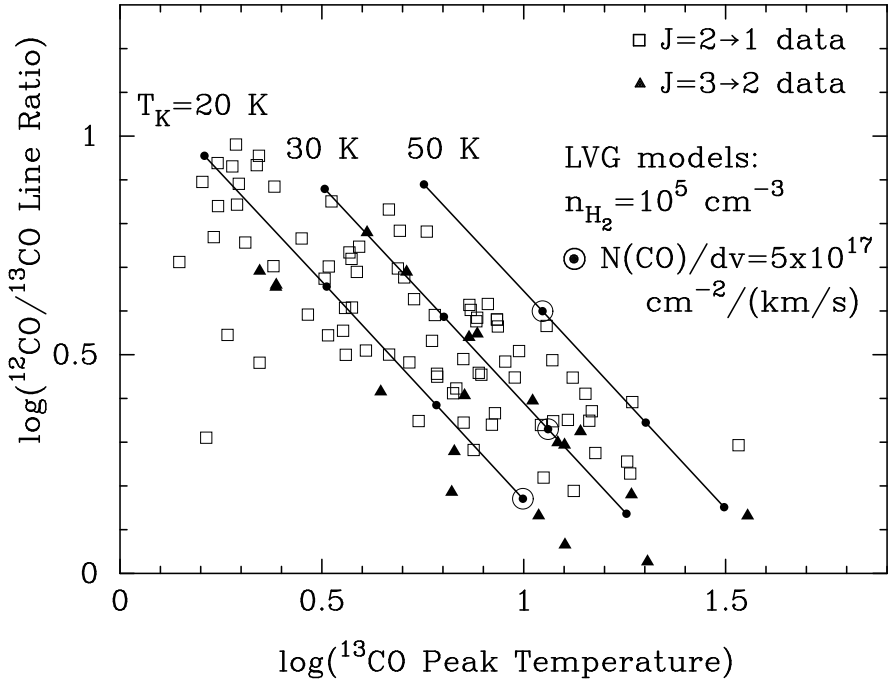}}
\caption{$^{12}$CO/$^{13}$CO line ratios versus $^{13}$CO peak temperature
in the Galactic molecular cloud M17 (Wilson et al. 1999). 
Larger line ratios at lower peak temperatures are produced by gas with
lower total column densities. The models refer to the J=2-1 transition.}
\end{figure}

These conclusions are supported by recent studies of giant molecular clouds
in the Milky Way. Carpenter, Snell, and Schloerb (1995) mapped a 200x200 pc
region of the Gem OB1 molecular complex in 
both $^{12}$CO and $^{13}$CO with the FCRAO 14~m telescope. They found
that the $^{12}$CO/$^{13}$CO intensity ratio is smallest in regions
with strong $^{12}$CO emission
(2-5), higher in low intensity regions (10-15), and highest in regions where
only $^{12}$CO emission is detected ($47\pm8$). Roughly 50\% of the $^{12}$CO
luminosity originates in regions without detectable $^{13}$CO emission.
Calculating the gas mass from the optically thin $^{13}$CO line where
it is detected, and the $^{12}$CO line plus an assumed optical depth
at other locations, the authors estimate that the 50\%
of the $^{12}$CO luminosity from low brightness regions 
accounts for only 19\% of the
total mass. Most of the molecular gas in the cloud complex lies in
relatively low column density regions ($< 10^{22}$ cm$^{-2}$).

Wilson, Howe, and Balogh (1999) have carried these studies to
higher rotational transitions with a 10x20 pc map of the M17 molecular
cloud. They also see larger $^{12}$CO/$^{13}$CO line ratios in
the J=2-1 and J=3-2 transitions in regions with
 lower $^{13}$CO peak temperatures, which correspond to regions
with lower column densities
(Figure~3). The overall trend
 seen in Figure 3 is primarily due to variations in
the gas column density from point to point in M17. The 
$^{12}$CO/$^{13}$CO line ratios averaged over the whole cloud are
$4.5\pm0.7$ for the J=2-1 lines and $3.7\pm0.9$ for the J=3-2 lines.
These line ratios are significantly smaller than typical extragalactic
ratios of 7-13 and 5-17, respectively, and show that the trend of
increasing $^{12}$CO/$^{13}$CO line ratios on larger spatial scales is
present for all three of the lowest rotational transitions. Again,
the most likely explanation for the higher line ratios in other galaxies
is a substantial contribution to the $^{12}$CO emission lines from
low column density molecular gas.

\section{Atomic Carbon Emission in M33}

\begin{figure}[htb]
\centerline{\psfig{figure=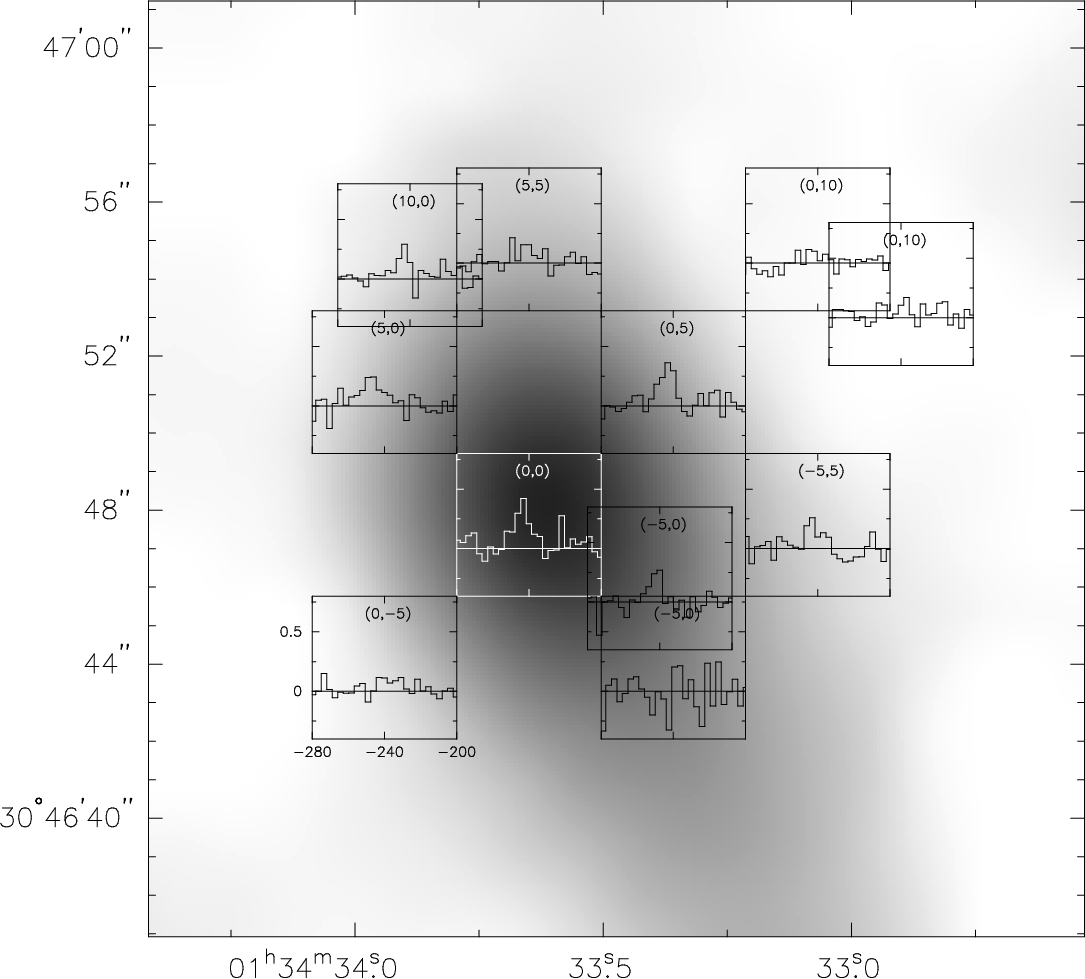,width=12truecm,%
       bbllx=32pt,bblly=84pt,bburx=564pt,bbury=709pt,angle=-90}}
\caption{The [CI] spectra from NGC 604-2 plotted on top of a greyscale image
of the CO J=1-0 emission from the Caltech Millimeter Array 
(Taylor and Wilson 2000). The [CI] beam is 12$^{\prime\prime}$. 
Notice the strong [CI] emission both towards
the peak of the CO emission and offest from the CO peak towards the northwest.}
\centerline{\psfig{figure=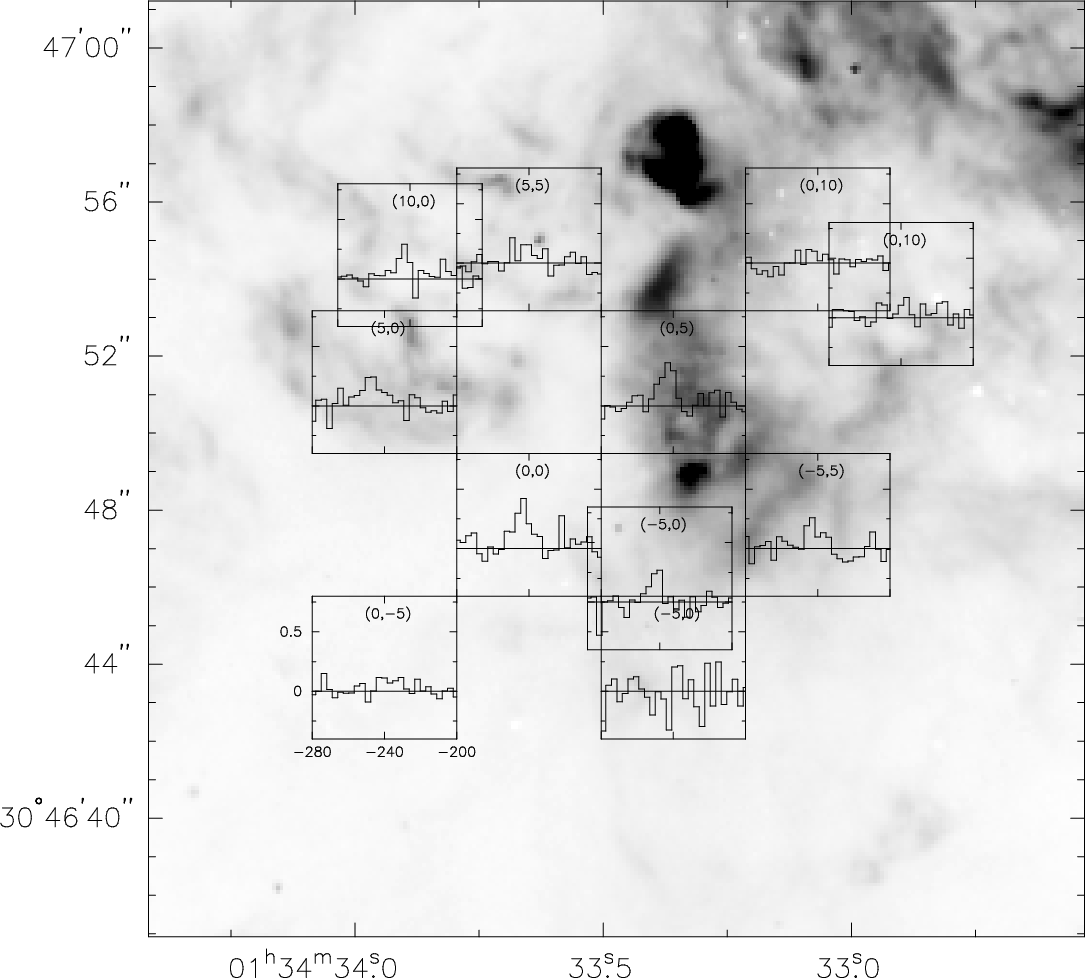,width=12truecm,%
       bbllx=32pt,bblly=84pt,bburx=564pt,bbury=709pt,angle=-90}}
\caption{The [CI] spectra from NGC 604-2 plotted on top of an H$\alpha$ image
from the HST archive (Taylor and Wilson 2000). The H$\alpha$ emission
in the northwest corner of the image is part of the giant $\HII$ region
NGC 604.}
\end{figure}

Atomic and ionized carbon provide important probes into the physical and
chemical structure of giant molecular clouds. Because the photodissociation
energy of CO and the ionization energy of C are very similar, atomic
carbon was originally expected to exist in only a thin layer sandwiched
between molecular and ionized regions. However, where
[CI] has been searched for over large enough areas, it is generally
found to be much more extended than expected (Keene et al. 1985;
Plume, Jaffe and Keene 1994; Plume et al. 1999). Since atomic carbon
can be produced by photodissociation of CO by ultraviolet radiation,
the primary interpretation of this large extent is that the molecular
clouds are sufficiently clumpy to allow radiation to penetrate deep
into the cloud. However, the large extent of the [CI] emission also raises
the question of whether nearby massive star formation is required to
produce atomic carbon, or whether it can be produced by the general
interstellar radiation field, perhaps via ultraviolet photons 
from field OB stars or leaking from $\HII$
regions (see also Hoopes and Walterbos, these proceedings). 
In addition, cosmic rays and 
chemical processes involving H$^+$ may also be able to produce
significant amounts of atomic carbon (Pineau des For\^ets, Roueff, and Flower
1992).

Mapping large regions of Galactic molecular clouds in [CI] emission
is time consuming even with specialized single-dish telescopes (i.e.
Sekimoto et al. 1999) or from space (i.e. Howe et al. 2000, and other
papers in the SWAS special issue). Nearby galaxies like M33 offer us
the opportunity to measure the total [CI] emission of individual
molecular clouds in just a few pointings. Wilson (1997) obtained the
first [CI] detections of four molecular clouds in M33 in single pointings
centered on the CO peak for each cloud. Atomic carbon was
detected in all four clouds, even in one cloud without an optical $\HII$
region. The presence of [CI] emission in this cloud suggests
that a nearby region of massive star formation is not required to 
produced atomic carbon.
The [CI] to CO integrated intensity ratio for the four clouds
shows significant variation, ranging from 0.04 to 0.18, with the two
clouds near optical $\HII$ regions showing the largest values.

One unusual result from the Wilson (1997) study is the relatively weak
[CI] emission from the molecular cloud NGC 604-2. Since this cloud is
located on the edge of the giant $\HII$ region NGC 604 and, in addition,
has below-solar metallicity, simple arguments suggested that atomic carbon
should be especially enhanced in this cloud. Taylor and Wilson (2000) 
obtained a small [CI] map of this cloud to determine the full extent of
the atomic carbon emission. Figure~4 shows the [CI] spectra overlaid
on a greyscale image of the CO J=1-0 emission obtained with the Caltech
Millimeter Array, while Figure~5 shows the [CI] emission compared to an
H$\alpha$ image. In addition to the [CI] line detected towards the peak
of the CO emission, there is an even stronger line located north-west
of the CO peak in the direction of  the $\HII$ region. The offset in the
[CI] emission towards the $\HII$ region suggests that atomic carbon
is created primarily by ultraviolet radiation from the nearby massive stars.
However, the presence of [CI] emission towards the center of the cloud
indicates a significant ionization fraction in the cloud's interior. 
However, [CI] emission is {\em not} detected towards the south-east side
of the cloud, even though the CO intensity is comparable to that of the 
north-west position where strong [CI] emission is seen. Overall, the asymmetry
in the [CI] emission suggests that the interior
ionization is most likely produced by penetration of the clumpy interior
of the cloud by ultraviolet radiation from the massive stars, rather than by
chemical processes or cosmic rays.

\section{Future Work}

With an impressive CO survey of the inner disk of M31 now
completed (Gu\'elin, these proceedings), the lack of a similar survey
for M33 now stands out as an important gap in the high-quality data sets
on the interstellar medium that are available for M31 and M33.
The large interferometric survey of the M33 disk currently
underway should result in a large increase in the number of molecular
clouds identified in this galaxy (Engargiola, Plambeck, and Blitz,
these proceedings). However, depending on how the sensitivity and 
filtering imposed by
the interferometric observations matches the mass and structural scale of the
molecular interstellar medium, this survey may not give us an accurate 
measure of the total molecular gas content of M33.
The only substantial extension to the early single dish
work of Wilson and Scoville (1989) is the recent major axis strip map of 
Corbelli (these proceedings), which reveals substantial CO emission
at several locations beyond the inner 1-2 kpc. 
The new high-resolution $\HI$ survey of
M33 (Thilker, these proceedings) offers the opportunity to
carry out an unbiased search for
CO emission associated with $\HI$ clouds. Such a survey would
likely identify molecular clouds across the entire disk, including clouds
in low-metallicity regions and regions with unusually low or high star
formation activity, and could be the starting point for future detailed
studies of the effect of metallicity and star formation 
on the molecular interstellar medium.
However, even such a targeted survey would not give us a complete measure
of the molecular gas content of M33. A complete CO survey of
the optical disk of M33 using FCRAO or an even larger telescope is
urgently needed. 

Another interesting area for future studies concerns the low mass end
of the cloud mass spectrum. The $^{13}$CO observations of M33 discussed
in Section 3 provide evidence for molecular gas at relatively low column
densities. This low column density material could exist in low mass clouds,
in more extended filamentary structures, or might even be simply the outer
regions of the more massive molecular clouds. Deep interferometric 
observations should be able to detect individual molecular clouds as
small as $10^4$ M$_\odot$, which may be the lower mass limit for
gravitationally bound clouds (Heyer, Carpenter, and Snell 2000; Heyer,
these proceedings). Identifying a population of low mass molecular clouds
would have important implications for understanding massive star formation
in M33. For example, Hoopes and Walterbos (these proceedings) have identified
a population of ``field'' O stars, i.e. O stars outside large OB
associations, which are responsible for 40\% of
the ionization of the diffuse interstellar gas. Where are the clouds
that could form these field O stars? A likely candidate would be a population
of $1-5\times 10^4$ M$_\odot$ molecular clouds, which would be
too faint to have been detected in the existing interferometric surveys.

In the broader picture, M33's proximity makes it an important testing ground
for the various proposed star formation laws (i.e. Kennicutt 1989) and for
testing models of molecular cloud formation and destruction. It is
interesting to see how clearly the two main spiral arms in M33 stand out
in the ISO 170 $\mu$m map (Hippelein, these proceedings); being able to
identify clearly arm and interarm regions is important for testing many
cloud formation scenarios. The relatively steep metallicity gradient in
M33 allows us to study the impact of metallicity on the molecular 
interstellar medium, from practical items such as the effect on the CO-to-H$_2$
conversion factor to more detailed physical properties such as the
distribution of carbon between the molecular, atomic, and ionized phases.
An important goal for the next decade is to understand in detail
the molecular interstellar medium in M33 and M31, both
the properties of the molecular gas itself 
and its relationship to other phases of the interstellar
medium and to star formation. This understanding will provide important
background to similar studies of more distant galaxies with a wider
range of physical and dynamical properties that
will be possible with the Atacama Large Millimeter Array.

\section*{References}\noindent

\references

Blitz L. \Journal{1985}{\ApJ}{296}{481}.

Carpenter J.M., Snell R.L., Schloerb F.P.
\Journal{1995}{\ApJ}{445}{246}.

Casoli F., Dupraz C., Combes F.
\Journal{1992}{\AaAp}{264}{55}.

Dame T.M., Koper E., Israel F.P., Thaddeus P.
\Journal{1993}{\ApJ}{418}{730}.

Encrenaz P.J., Stark A.A., Combes F., Wilson R.W.
\Journal{1979}{\AaAp}{78}{1}.

Heyer M.H., Carpenter J.M., Snell R.L.
\Journal{2000}{\ApJ}{}{}in preparation.

Howe J.E. et al.
\Journal{2000}{\ApJL}{}{}in press.

Keene J., Blake G.A., Phillips T.G., Huggins P.J., Beichman C.A.
\Journal{1985}{\ApJ}{299}{967}.

Kennicutt R.C.
\Journal{1989}{\ApJ}{344}{685}.

Lada E.A., DePoy D., Evans N.J., Gatley I. 
\Journal{1991}{\ApJ}{371}{171}.

Loinard L., Dame T.M., Heyer M.H., Lequeux J., Thaddeus P.
\Journal{1993}{\AaAp}{351}{1087}.

Neininger N., Gu\'elin M., Ungerechts H., Lucas R., Wielebinski R.
\Journal{1998}{\Nat}{395}{871}.

Pineau des For\^ets G., Roueff E., Flower D.R.
\Journal{1992}{\MNRAS}{258}{P45}.

Plume R., Jaffe D.T., Keene J.
\Journal{1994}{\ApJL}{425}{49}.

Plume R., Jaffe D.T., Tatematsu K., Evans N.J., Keene J.
\Journal{1999}{\ApJ}{512}{768}.

Polk K.S., Knapp G.R., Stark A.A, Wilson R.W.
\Journal{1988}{\ApJ}{332}{432}.

Sekimoto Y. et al. (1999) in {\em Proceedings of Star Formation 1999},
ed. T. Nakamoto, Nobeyama Radio Observatory, p.~86.

Taylor C.L., Wilson C.D.
\Journal{2000}{\ApJ}{}{}in press.

Turner B.E. \Journal{1984}{Vistas Astron.}{27}{303}.

Viallefond F., Boulanger F., Cox P., Lequeux J., P\'erault M., Vogel S.N.
\Journal{1992}{\AaAp}{265}{437}.

Wilson C.D. \Journal{1997}{\ApJL}{487}{49}.

Wilson C.D., Howe J.E., Balogh M.L. \Journal{1999}{\ApJ}{517}{174}.

Wilson C.D., Scoville N. \Journal{1989}{\ApJ}{347}{743}.

Wilson C.D., Scoville N. \Journal{1990}{\ApJ}{363}{435}.

Wilson C.D., Scoville N. \Journal{1992}{\ApJ}{385}{512}.

Wilson C.D., Scoville N.Z., Freedman W.L., Madore B.F., Sanders D.B.
\Journal{1988}{\ApJ}{333}{611}.

Wilson C.D., Walker C.E. \Journal{1994}{\ApJ}{432}{148}.

Wilson C.D., Walker C.E., Thornley M.D. \Journal{1997}{\ApJ}{483}{210}.

Young J.S., Scoville N. \Journal{1982}{\ApJL}{260}{11}.

\end{document}